# High frequency sheath modulation and higher harmonic generation in a low pressure very high frequency capacitively coupled plasma excited by sawtooth waveform

Sarveshwar Sharma[1,2], Nishant Sirse[3] and Miles M Turner[4]

[1]Institute for Plasma Research, Gandhinagar – 382428, India

[2]Homi Bhabha National Institute, Anushaktinagar, Mumbai – 400094, India

[3]Institute of Science & Laboratory Education, IPS Academy, Indore -452012, India

[4]School of Physical Sciences and National Center for Plasma Science and Technology, Dublin City University, Dublin 9, Ireland

E-mail: nishantsirse@ipsacademy.org

**Abstract**

A particle-in-cell (PIC) simulation study is performed to investigate the discharge asymmetry, higher harmonic generations and electron heating mechanism in a low pressure very high frequency capacitively coupled plasma (CCP) excited by a saw-tooth like current waveform. Two current densities, 50 A/m$^2$ and 100 A/m$^2$ are chosen for a constant gas pressure of 5 mTorr in argon plasma. The driving frequency is varied from 13.56 MHz to 54.24 MHz. At a lower driving frequency, high frequency modulations on the instantaneous sheath edge position at the grounded electrode are observed. These high frequency oscillations create multiple ionization beam like structures near to the sheath edge that drives the plasma density in the discharge and responsible for discharge/ionization asymmetry at lower driving frequency. Conversely, the electrode voltage shows higher harmonics generation at higher driving frequencies and corresponding electric field transients are observed into the bulk plasma. At lower driving frequency, the electron heating is maximum near to the sheath edge followed by electron cooling within plasma bulk, however, alternate heating and cooling i.e. burst like structures are obtained at higher driving frequencies. These results suggest that electron heating in these discharges will not be described accurately by simple analytical models.

## 1. Introduction

Capacitively coupled plasmas (CCPs) are one of the most fundamental etching tools used in the semiconductor industries for the fabrication of large-scale integrated circuits [1]. An independent control of ion flux and mean ion energy is vital for plasma etching and it was first made achievable by CCP discharge excited with help of dual-frequency (DF) by Goto et al. [2]. The high-frequency voltage amplitude largely controls the plasma production i.e. charge particle densities and the ion energy is determined by the low-frequency voltage amplitude as ions transit across the sheath. The CCP discharges driven by DF waveforms under different operating conditions by using simulation and experiments have thoroughly been studied in the literature [3-8]. The studies also reported that frequency coupling effects and secondary electron emission from the electrodes can constraint the independent control of ion properties in DF CCPs [9-10]. In one of the research findings, it was reported that *ion energy distribution function* (IEDF) can be controlled by imposing an intermediate frequency in DF-CCPs [11]. In this type of triple frequency excited CCP configuration, the effect of intermediate frequency on plasma parameters has been studied by various numerical simulation techniques [11-15]. The other novel technique, Electrical Asymmetric Effect (EAE) was proposed to control the ion properties, introduced by Heil et al [16-17]. In this technique, the plasma was excited with the help of a fundamental frequency and its second harmonic that lead to the formation of a DC self-bias even in geometrically symmetrical CCP discharges. The phase between the two harmonics can control the self-bias that can directly affect the energy of the ions at the electrodes without affecting the ion flux [18]. The implementation of the EAE has been thoroughly examined in both electropositive and electronegative plasmas comprising different physical mechanisms like power absorption by electrons, effect of the secondary electron emission and driving frequencies [19-22].

In a recent development, the excitation of capacitively coupled radio-frequency discharges with tailored waveforms has emerged as a promising tool for an effective control of ion flux and ion energy in technological plasmas [23]. These tailored waveforms are also known as customized waveforms which have non-sinusoidal shape [24-30] and can be generated by superimposition of signal with a fundamental frequency and its higher harmonics with an appropriate phase shift between them. The shape of the waveform can be modified by changing the harmonics amplitude and their phases. In recent years, tailored waveforms played an important role in silicon etching [31-32] and thin film deposition [33-34]. More specifically, a novel type of temporally asymmetric waveform known as sawtooth waveform, which is the family of tailored waveform introduced by Bruneau et al [35-37]. It consists of a fast rise and slow fall or vice versa that has slope asymmetry but no amplitude asymmetry. The amplitude asymmetry has been comprehensively examined for dual-frequency excitation [38-41] and for the waveforms including more harmonics [42-47]. In the slope asymmetry sawtooth-like waveforms are used, which leads to significantly different sheath dynamics [35-36]. For voltage driven CCP, Bruneau et al. experimentally shown by using phase resolved optical emission spectroscopy (PROES) in an argon plasma that there is a strong ionization close to the electrode where sheath expansion is fast and weak ionization at other electrode where sheath expansion is slow for sawtooth waveform [36].

The work of Bruneau et al. [35] reported the effect of sawtooth voltage waveform with different rise and fall slopes. They performed PIC simulation in argon gas with electrode gap of 2.5 cm, 13.56 MHz applied frequency, neutral gas pressure of 400 mTorr and peak to peak applied voltage 200 V. Due to high gas pressure, the secondary electron emission was considered in their study. Their results showed that the fast sheath expansion adjacent to powered electrode triggers an enormous pressure heating, which results a high ionization rate near to the sheath edge. On the other hand, gradual sheath expansion near to the grounded electrode results in weak heating and a low ionization rate. Subsequently, a large ion flux ($\Gamma_i$) asymmetry develops with a lower flux reaching on the grounded electrode. They also observed that the appearance of significant positive self-bias when sawtooth waveform applied. So, this research work reports the effects of slope asymmetry using that one can independently control the ion flux on each electrode. Later in 2015, Bruneau et al. [36] expanded their previous work for the same set of simulation parameters as described above. The effect of pressure (20 mTorr to 800 mTorr) on discharge asymmetry was investigated at 200 V discharge voltage and 13.56 MHz applied frequency. The simulation outcome showed that at low pressure, the ionization mainly occurs in the bulk plasma and no asymmetry was seen. The discharge shifts from bulk ionization to the sheath ionization as pressure increases and the asymmetry appears. As the pressure further increased the asymmetry is more intense because of the significant local ionization near to the sheath edge. Another important study reported in this paper is the effect of changing the base frequency (1.695 MHz, 3.39 MHz, 13.56 MHz and 54.24 MHz) at 400 mTorr gas pressure. It was described that the time averaged ionization rate ($<K_{iz}>$) is strongly reduced when applied frequency ($f_{rf}$) is reduced i.e. it is maximum at 54.24 MHz and minimum at 1.695 MHz. It is noted that the maximum of $<K_{iz}>$ moves away from the powered electrode when $f_{rf}$ is decreased because of the larger sheath width. Their explanation was based on two opposing mechanisms. With decreasing $f_{rf}$, the asymmetric effect is lower and the ionization peak shifts away from the powered electrode because diffusion to the opposing electrode is facilitated. On the other side, when the local maxima close to the grounded electrode is lower the asymmetry is higher because the asymmetry of ion creation is increased. As a result, when $f_{rf}$ is reduced two mechanisms work, first one which increases the asymmetry and the other one which tends to decrease it. Therefore, the asymmetry behavior depends on the dominant effect. It was reported that density is strongly reduced when $f_{rf}$ is reduced from 54.24 to 3.39 MHz and the density profile is more and more asymmetric. This explains that the second mechanism as described above, i.e. the drop of ionization peak near to grounded electrode, plays the dominant role. However, the applied frequency effect at a low gas pressure i.e. in the collisionless regime, which is typically the case in plasma processing was not reported in the literature. In the present manuscript, we show the effect of driving frequency in a nearly collisionless regime. The asymmetry effect is investigated along with the higher harmonic generations and electron heating mechanism. The simulation is performed for a perfect saw-tooth current waveform that contains a large number of harmonics.

The article is structured as follows. The simulation technique that is established on Particle-in-Cell/Monte Carlo collision (PIC/MCC) methods is described in section II. The physical interpretation and description of simulation results are presented in section III. In section IV, the summary and conclusion are given.

## 2. Simulation Technique and Parameters

A *current driven* symmetric capacitively coupled discharge in argon plasma is simulated here by using a very well tested and benchmarked 1D3V, electrostatic, self-consistent, particle-in-cell (PIC) code. The Particle-in-Cell/Monte Carlo collision (PIC/MCC) methods is the core of this simulation technique and its detail can be found in literature [48-49]. This code is developed at Dublin City University by Prof. Miles Turner and extensively used in several research articles and few of them can be find here [50-60]. This code can run in both current and voltage driven modes and we have opted for the current driven mode for the present paper. The details of the simulation technique are reported in literature [61-62]. In our simulation, we have considered all important particle particle collision reactions like ion-neutral (elastic, inelastic and charge exchange), electron-neutral (elastic, inelastic and ionization) and processes like multi-step ionization, metastable pooling, partial de-excitation, super elastic collisions and further de-excitation. The type of reactions and others details are given in the literature [57]. The creation of metastables (i.e. Ar*, Ar**) has been considered in simulation and also tracked them in our output diagnostics. These two lumped excited states of Ar, i.e., Ar* ($3p^54s$), 11.6 eV, and Ar** ($3p^54p$), 13.1 eV, in uniform neutral argon gas background is considered with charged particles viz. electrons and ions in simulation. The details of species included in simulation, their reactions and cross sections are taken from well-tested sources in the literature [52, 57, 63]. The stability and accuracy criterion of PIC is addressed by choosing appropriate grid size ($\Delta x$) and time step size ($\Delta t$) that can resolve the Debye length and the electron plasma frequency respectively. It is assumed that both the electrodes have infinite dimension and are planar and parallel to each other. We have also considered that electrodes are perfectly absorbing and for the sake of simplicity secondary electron emission is ignored. The discharge gap is 6 cm and is operated at a gas pressure of 5 mTorr. The gas is uniformly distributed with a fixed temperature 300 K. The temperature of ions is the same as the neutral gas. The number of particles per cell is 100 for all sets of simulation.

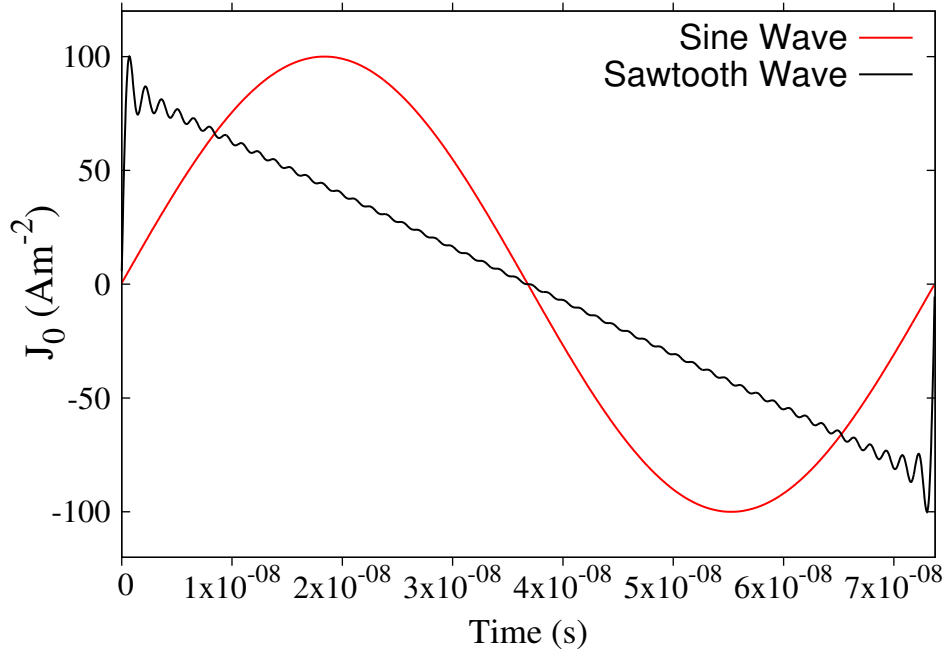

**Figure 1:** Profile of sinusoidal and sawtooth current waveform for 13.56 MHz having current density amplitude of 100 A/m$^2$.

The complex structures of applied waveform can be produced by employing an adequately high number of harmonics of a fundamental RF frequency and sensibly chose phase shifts between them. For current driven CCP, the sawtooth waveform has following expression (see figure 1):

$$J_{rf}(t) = \pm J_0 \sum_{k=1}^{N} \frac{1}{k} \sin(k\omega_{rf} t) \quad \text{----------(1)}$$

where positive and negative signs correspond to "sawtooth-down" and "sawtooth-up" waveforms respectively. Here, $J_o$ is the current density amplitude applied at the powered electrode, and $\omega_{rf}$ is the fundamental angular driven radiofrequency. The magnitude $J_0$ varies with the total number of harmonics i.e. N (which is 50 here) and is arranged in a way to provide the required peak-to-peak current density amplitude. We have used "sawtooth-down" current waveform for all set of our simulations.

## 3. Results and Discussions

The velocity of energetic electrons produced from near to the sheath edge is affected by the shape of applied waveform and driving frequency i.e. it influences the sheath dynamics significantly and therefore the plasma profile changes drastically. Figure 2 (a) and 2 (b) shows the time average ion and electron density profile within discharge system at 13.56 MHz, 27.12 MHz and 54.24 MHz driving frequency for current density amplitudes of 50 A/m$^2$ and 100 A/m$^2$ respectively. In figure 2, the RF current source is applied at 0 cm and the grounded electrode is at 6 cm. Strong asymmetry is observed at lower driving frequencies. At 13.56 MHz and 50 A/m$^2$, the results showed that the sheath width near to the grounded electrode is broader (~ 10 mm) when compared to the sheath width near to powered electrode (~ 6 mm). The sheath width is estimated as where the electron sheath edge is at maximum distance from the electrode and quasi-neutrality breaks down. A sharp dip in the electron density is observed at this point when moving from bulk plasma towards electrode. When the applied frequency is increased from 13.56 MHz to 27.12 MHz, the sheath width near grounded and powered electrode are ~ 5 mm and ~4 mm respectively. The density profile is more symmetric compared to 13.56 MHz case. When the frequency is further increased to 54.24 MHz, the density profile is nearly symmetric and the sheath width near to the electrodes are approximately 3 mm. Similar trend is observed at a current density of 100 A/m$^2$. Above description demonstrates that sheath width can be controlled by changing the applied frequency in sawtooth waveform case hence the ion energy can be effectively controlled.

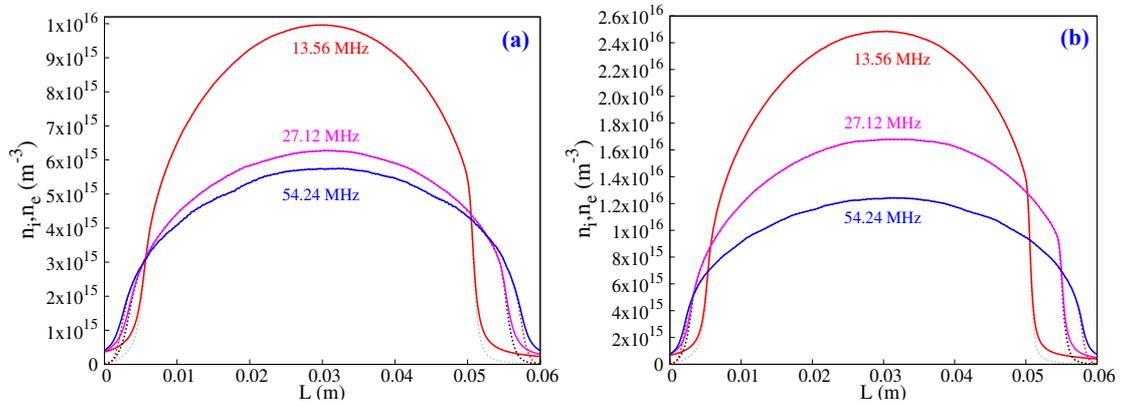

**Figure 2:** The profile of time average ion (solid line) and electron density (dotted line) at different applied frequencies i.e. 13.56 MHz, 27.12 MHz and 54.24 MHz for current densities of (a) 50 A/m$^2$ and (b) 100 A/m$^2$.

With increasing driving frequency, the symmetric behavior of the discharge agrees with the results obtained by Bruneau et al [36] for a high-pressure voltage driven case. However, in their study the density shows the opposite trend i.e. it was increasing with driving frequency. In the present current driven case, the density is decreasing with an increase in driving frequency. At 50 A/m$^2$, the central peak density is 9.96x10$^{15}$ m$^{-3}$ at 13.56 MHz and decreasing to 5.7x10$^{15}$ m$^{-3}$ at 54.24 MHz. Similarly, at 100 A/m$^2$, the central peak density is 2.5x10$^{16}$ m$^{-3}$ at 13.56 MHz and decreasing to 1.24x10$^{16}$ m$^{-3}$ at 54.24 MHz i.e. decrease by approximately 50%. Similar kind of decreasing trend is observed in spatially averaged densities. The corresponding bulk electron temperature for 50 A/m$^2$ is increasing from 1.68 eV at 13.56 MHz to 2.3 eV at 54.24 MHz, and for 100 A/m$^2$ it is increasing from 1.65 eV at 13.56 MHz to 2.35 eV at 54.24 MHz. At a higher driving frequency, the symmetric behavior of the discharge and a decrease in plasma density can be explained by analyzing the ionization rate within the discharge system. Figure 3 (i, a) and 3 (ii, a) shows the time averaged ionization rate for different driving frequencies at 50 A/m$^2$ and 100 A/m$^2$ current density amplitudes respectively. The corresponding time averaged excitation rates for different excited species are shown in figures 3 (b) to 3 (d). As shown in figure 3, due to longer electron mean free path at lower gas pressure, the ionization and different excitation mechanisms are mostly occurring into the bulk plasma. The ionization rate (figure 3 (i, a) and 3 (ii, a)) is higher at 13.56 MHz and decreasing at 54.24 MHz. The corresponding excitation rates to Ar* (3p$^5$4s), 11.6 eV (figure 3(i, b) and figure 3(ii, b)), and Ar** (3p$^5$4p), 13.1 eV (figure 3(i, c) and figure 3(ii, c)) is also higher at 13.56 MHz and decreasing with a rise in driving frequency. The low energy electrons (~1.5 eV) are further consumed in excitation from Ar* to Ar**, and its rate is highest at 13.56 MHz and decreasing drastically at 54.24 MHz (figure 3(i, d) and figure 3(ii, d)), which may be attributed to the presence of high energy electrons generated from near to the sheath edge and increase the bulk electron temperature. Thus the population of low energy electrons responsible for further excitation are reduced. It is also noticed that, at a higher driving frequency, the ionization rate and different excitation rates are extending further close to the electrodes, i.e. becomes symmetric, bulk plasma length increases and therefore responsible for the symmetric behavior of the discharge as observed in the plasma density profile shown in figure 2.

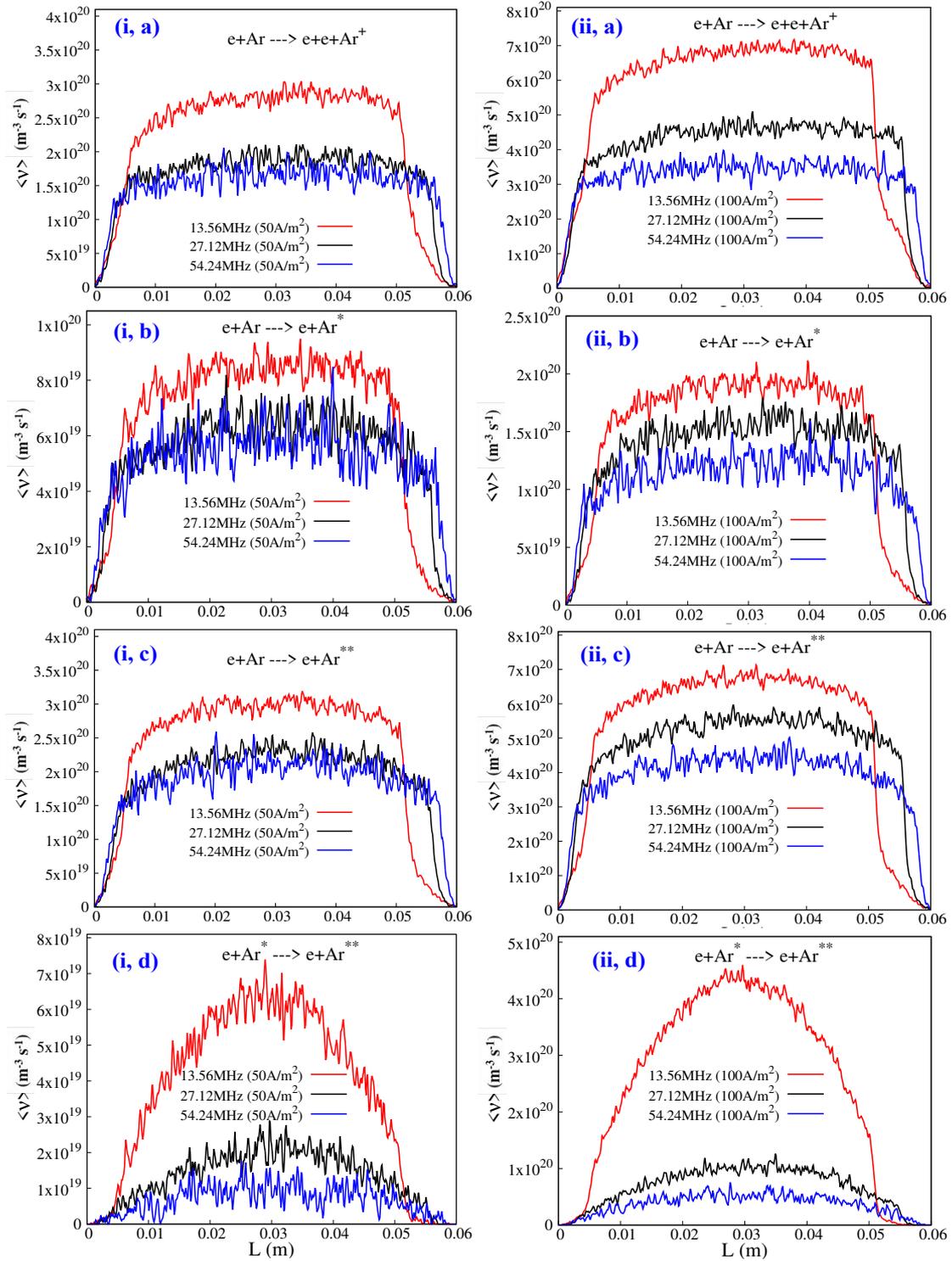

**Figure 3:** Time averaged ionization and excitation rates at different applied frequencies i.e. 13.56 MHz, 27.12 MHz and 54.24 MHz for current densities of 50 A/m$^2$ and 100 A/m$^2$.

To investigate further the ionization mechanism, the spatio-temporal evolutions of ionizing collision rate for different driving frequencies are examined. This is shown in figure 4. The spatio-temporal plots are averaged over last 600 RF cycles in steady state. At 50 A/m$^2$ and 13.56 MHz driving frequency (figure 4 (i, a)), a strong asymmetry is observed in the ionization collision rate i.e. it is higher near to grounded electrode when compared to powered electrode and highest during the expanding phase of the sheath. Furthermore, due to nearly collisionless operation the ionization front is penetrating into the bulk plasma and extending up to the opposite sheath. As driving frequency increases to 27.12 MHz, the ionization collision rate asymmetry reduces, and at 54.24 MHz driving frequency it is nearly symmetric. Comparing different driving frequency for constant current density (figure 3 (i, a), 3 (i, b) and 3 (i, c) or figure 3 (ii, a), 3 (ii, b) and 3 (ii, c)), the ionization collision rate is highest for 13.56 MHz driving frequency. Comparing different current densities, the ionization collision rate is higher at 100 A/m$^2$ current density when compared to 50 A/m$^2$, however, the effect of increasing driving frequency is similar for both current densities.

At 13.56 MHz, the asymmetric behavior of the ionization rate is attributed to the enhanced pressure (collisionless) heating near to the grounded electrode. As described by the hard wall model [64], one of the factors responsible for enhanced collisionless heating is the sheath velocity. In the present case, we observed that the sheath width is decreasing, however, the sheath velocity is increasing with an increase in driving frequency. This will increase the energy gained by the electrons, and hence ionization collision rate and plasma density should increase. However, we observe an opposite trend i.e. the plasma density/ionizing collision rate is decreasing with driving frequency. This suggests that a different phenomenon is responsible here. Further inspecting the spatio-temporal ionization rate, it is noticed that at 13.56 MHz multiple beams like ionization collision rate is present near to the expanding sheath edge. The number of multiple ionization beams are higher at higher current density and decrease with driving frequency for both current densities. These multiple beams are caused due to high frequency modulation on the instantaneous sheath edge position analogue to multiple frequency excited CCP discharges. We believe that the generation of multiple beams from near to the expanding sheath edge drive the higher ionization rate/plasma density in the discharge. In the following discussion, we further investigate the non-linearity in electric field and electrode voltage to support above fact.

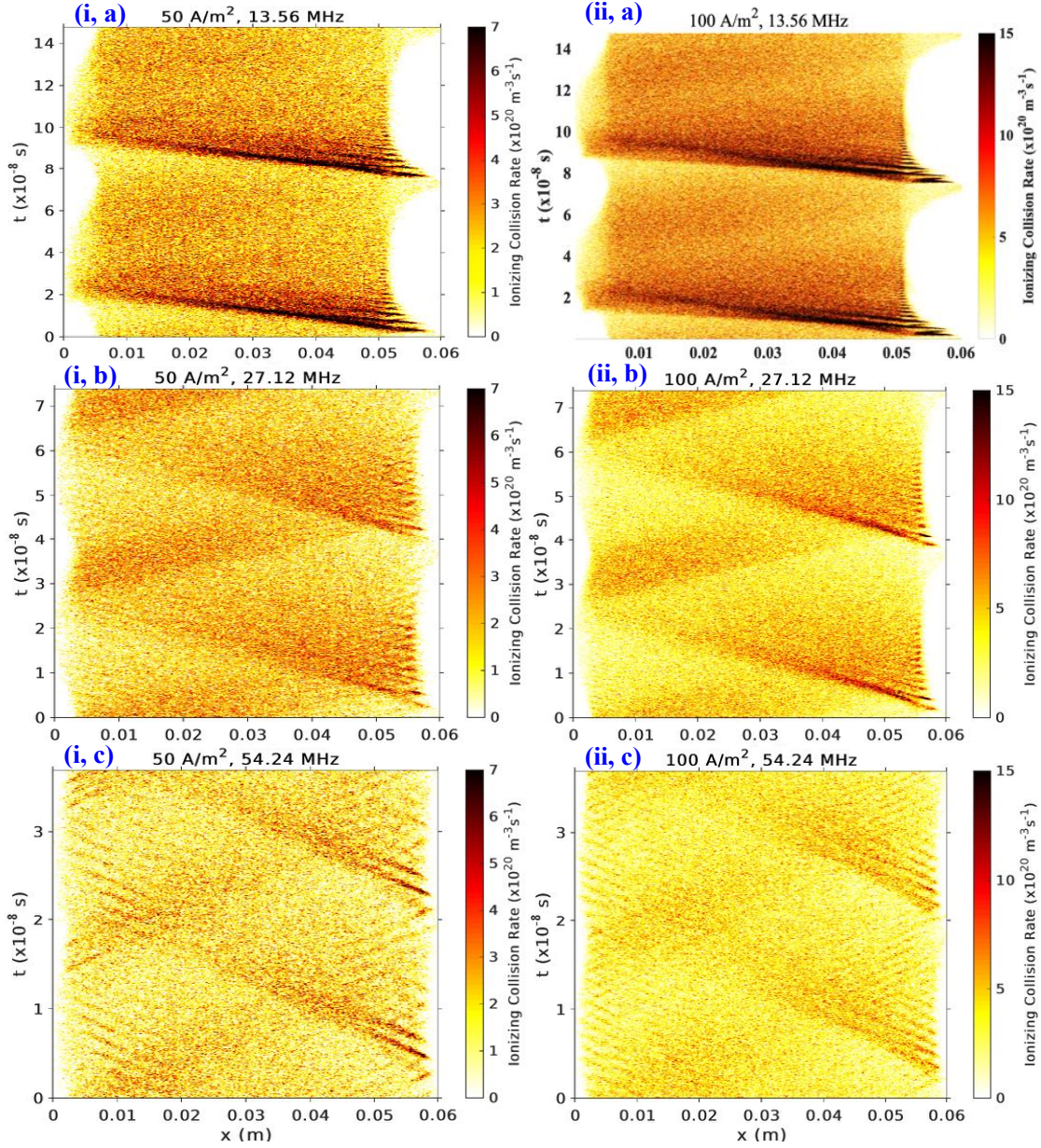

**Figure 4:** Spatio-temporal evolution of ionizing collision rates at different applied frequencies i.e. 13.56 MHz, 27.12 MHz and 54.24 MHz for current densities of 50 A/m$^2$ and 100 A/m$^2$.

To discuss the higher harmonics generation in the discharge, we first investigate the spatio-temporal evolution of electric field in the discharge. This is plotted in figure 5 for different driving frequencies and for current densities of 50 A/m$^2$ and 100 A/m$^2$. The spatio-temporal plots are averaged over last 100 RF cycles in steady state. As shown in figure 5 (i, a), at 13.56 MHz, the electric field is mostly confined into the sheath region and bulk plasma is almost quasi-neutral. However, the instantaneous sheath edge position near to the grounded electrode is significantly modified i.e. high frequency oscillations are observed on the instantaneous sheath edge position. When the driving frequency increase to 27.12 MHz (figure 5 (i, b)), the electric field transients appears into the bulk plasma. The corresponding high frequency oscillation on the instantaneous sheath edge is reduced. At 54.24 MHz (figure 5 (i, c)), the electric field transients are strongest, reaching up to the opposite sheath edge and weakly modifying its instantaneous position. The high frequency oscillations on the grounded instantaneous sheath edge position is substantially reduced. As explained earlier, the high frequency modulation drive ionization rate/plasma density in the discharge. A reduction in the frequency of sheath modulation explains why the ionization rate and plasma density is decreasing with a rise in driving frequency. On the other hand, the energy of the beam is driven by the sheath velocity, which is increasing with driving frequency and therefore one can observe high energy electrons produced at 54.24 MHz reaching up to the opposite sheath edge. This is consistent at higher current density i.e. at 100 A/m$^2$. As observed in figure 5 (ii, a), (ii, b) and (ii, c), the high frequency modulation on the instantaneous sheath edge position is decreasing with an increase in driving frequency. When comparing the same driving frequencies, for two different current densities, one can observe that the ionization rate/plasma density (figure 3/ figure 2) is higher at higher current density due to enhanced high frequency oscillations on the instantaneous sheath edge position. The corresponding beam energy is also higher at higher current density due to greater sheath width/sheath velocity. This could be appropriately envisaged at higher driving frequency i.e. at 54.24 MHz (figure 5 (i, c) and figure 5 (ii, c)), where the high frequency oscillations are higher at 100 A/m$^2$ (figure 5 (ii, c)) when compared to 50 A/m$^2$ (figure 5 (i, c)). The corresponding beam energy is also higher at 100 A/m$^2$ (figure 5 (ii, c)) that is reaching up to the opposite sheath edge and generating the high frequency modulation on the powered electrode sheath. However, the beam energy is lower at 50 A/m$^2$ (figure 5 (i, c)) i.e. reaching up to the opposite sheath edge but weakly modifying it. The above results conclude that the high frequency modulation drive plasma density and the sheath velocity drive beam energy. At 27.12 MHz (figure 5 (i, b) and (i, c)) and 54.24 MHz (figure 5 (ii, b) and (ii, c)), it is further noticed that the electric field transients are not continuous i.e. burst like structures are observed. By observing the time-averaged electron heating we will show that this will drive the alternate electron heating and cooling in the bulk plasma.

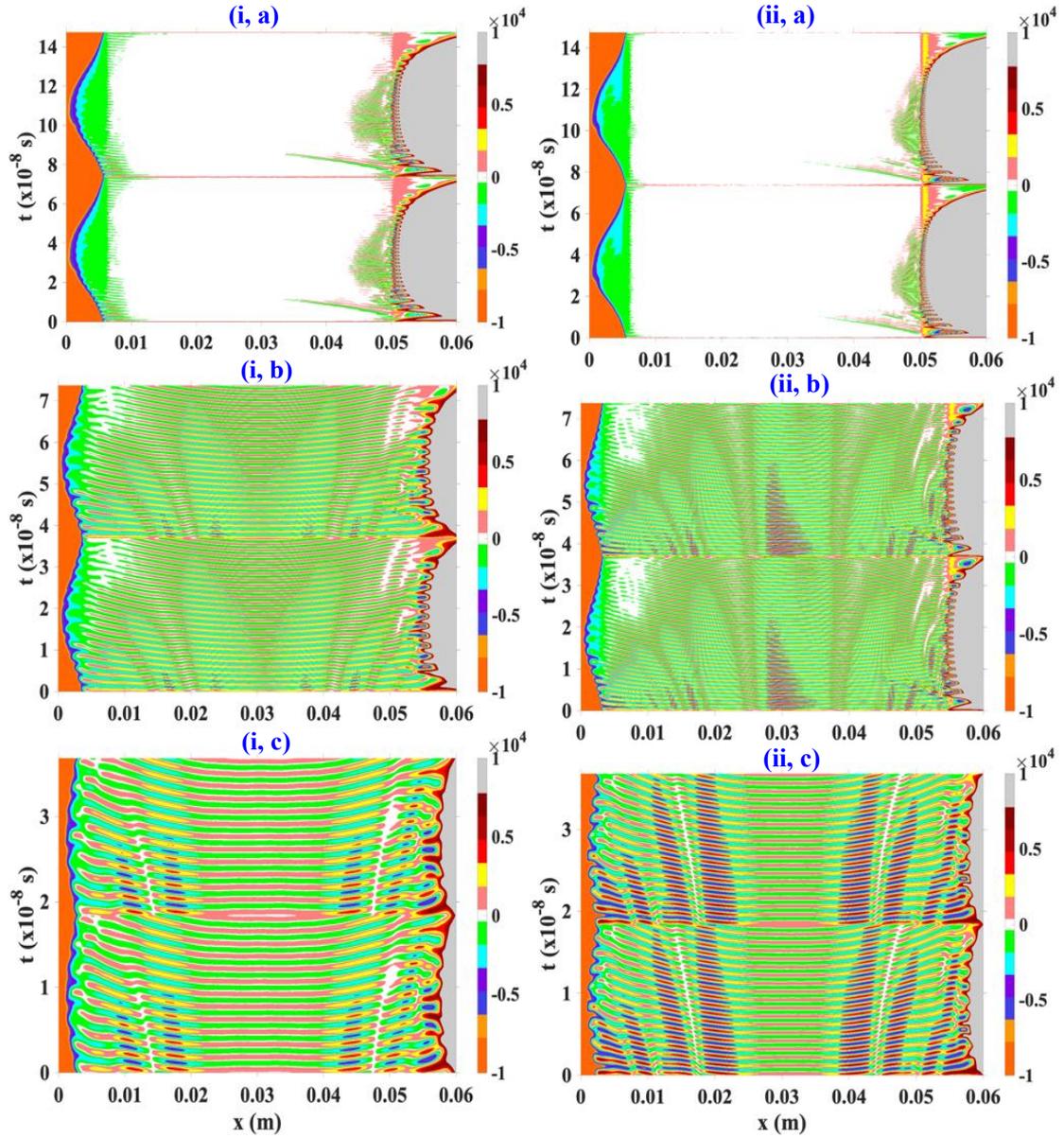

**Figure 5:** Spatio-temporal evolution of electric field at different applied frequencies for 50 A/m$^2$ and 100 A/m$^2$: (i, a) 13.56 MHz (50 A/m$^2$), (i, b) 27.12 MHz (50 A/m$^2$), (i, c) 54.24 MHz (50 A/m$^2$), (ii, a) 13.56 MHz (100 A/m$^2$), (ii, b) 27.12 MHz (100 A/m$^2$), and (ii, c) 54.24 MHz (100 A/m$^2$).

The frequency of instantaneous sheath edge modulation could be inferred by transforming the simulation results from time domain to frequency domain. For this purpose, we plot the electric field at the centre of the discharge and electrode voltage along with its fast fourier transform (FFT). This is shown in figure 6 and figure 7. As shown in figure 6 (i, a), the dominant higher harmonic of frequency 68 (~900 MHz) times the fundamental is triggered into the bulk plasma. The amplitude of all the harmonics are below 8%. The triggered dominant harmonic is slightly over electron plasma frequency at the centre of the discharge, which is ~898 MHz in this case. As driving frequency increase to 27.12 MHz (figure 6 (i, b)), the frequency of sheath modulation decrease, a distinct 26$^{th}$ (~705 MHz) harmonic (dominant frequency) along with its nearby harmonics are penetrating the bulk plasma. The contribution of this dominant higher harmonic is higher in comparison to 13.56 MHz (figure 6 (i, a)). The triggered frequency is slightly higher than electron plasma frequency (~700 MHz) at the centre of the discharge. At 54.24 MHz (figure 6 (i, c)), the frequency of sheath modulation is further reduced to 650 MHz (12$^{th}$ harmonic) and several other harmonics are observed (labelled in the figure). The electron plasma frequency is ~650 MHz in this case. At this driving frequency, the amplitude is now distributed over large number of higher harmonics and therefore the amplitude of dominant harmonic (12$^{th}$ harmonics) is lower than the amplitude of dominant harmonic (26$^{th}$ harmonic) observed at 27.12 MHz. Similar results are observed at 100 A/m$^2$ (figure 6 (ii, a), (ii, b) and (ii, c)) i.e. the frequency of dominant harmonic is decreasing with driving frequency and several numbers of higher harmonics are observed at higher driving frequency.

Figure 7 shows the corresponding electrode voltage and its FFT. The investigation of electrode voltage is useful not only for the further explanation of simulation outcomes but also could be measured experimentally. We first notice that at both current densities the amplitude of electrode voltage decreases with an increase in driving frequency, from ~865 V to ~190 V at 50 A/m$^2$ and ~1790 V to ~290 V at 100 A/m$^2$. For same driving frequency, the electrode voltage is higher at 100 A/m$^2$ when compared to 50 A/m$^2$ current density. This is attributed to a decrease in the sheath width with driving frequency, which allows higher current to flow through the discharge and therefore electrode voltage drops. We also notice that, at higher driving frequencies, higher harmonics are observed in the electrode voltage that are similar to ones appeared in electric field transients (figure 6). However, the contribution/amplitude of higher harmonics are comparatively lower when compared to higher harmonics in the electric field at the centre of the discharge. At 50 A/m$^2$ and 13.56 MHz (figure 7 (i, a)), the FFT shows only fundamental frequency and no higher harmonics are observed. This is due to no electric field penetration (figure 5 (i, a)) into the bulk plasma at this operating condition. The corresponding electric field FFT at the centre of the discharge (figure 6 (i, a)) also shows weak presence of higher harmonics. As driving frequency increase to 27.12 MHz (figure 7 (i, b)), the fundamental frequency drops and 26$^{th}$ (~705 MHz) harmonic appears in the FFT. This is similar to the one observed in electric field transient (figure 6 (i, b)). Further increase in driving frequency to 54.24 MHz (figure 7 (i, c)), the fundamental frequency reduces to below 20% and several higher harmonics are observed with a dominant one at 12$^{th}$ harmonic (~650 MHz), which is similar to that observed in electric field transients (figure 6 (i, c)). These higher harmonics are appeared due to non-linear interaction of electrons with the oscillating sheath edge responsible for modifying the instantaneous sheath edge positions and generating the multiple electron beams. In contrast to sinusoidal case [65], a current driven CCP excited by saw-tooth waveform could generate higher harmonics in the electrode voltage due to several harmonics composed in the applied current waveform. In the previous study of sinusoidal case [65], it was noticed that a strong electric field reversal is created at an instant after the beam acceleration and thereafter attracts bulk electrons for further acceleration. For saw-tooth like waveform, the sheath edge expansion is much faster in comparison to sinusoidal case and therefore one could expect greater energy transfer and therefore higher energetic electrons acceleration. A detailed investigation of this mechanism is out of scope of this work.

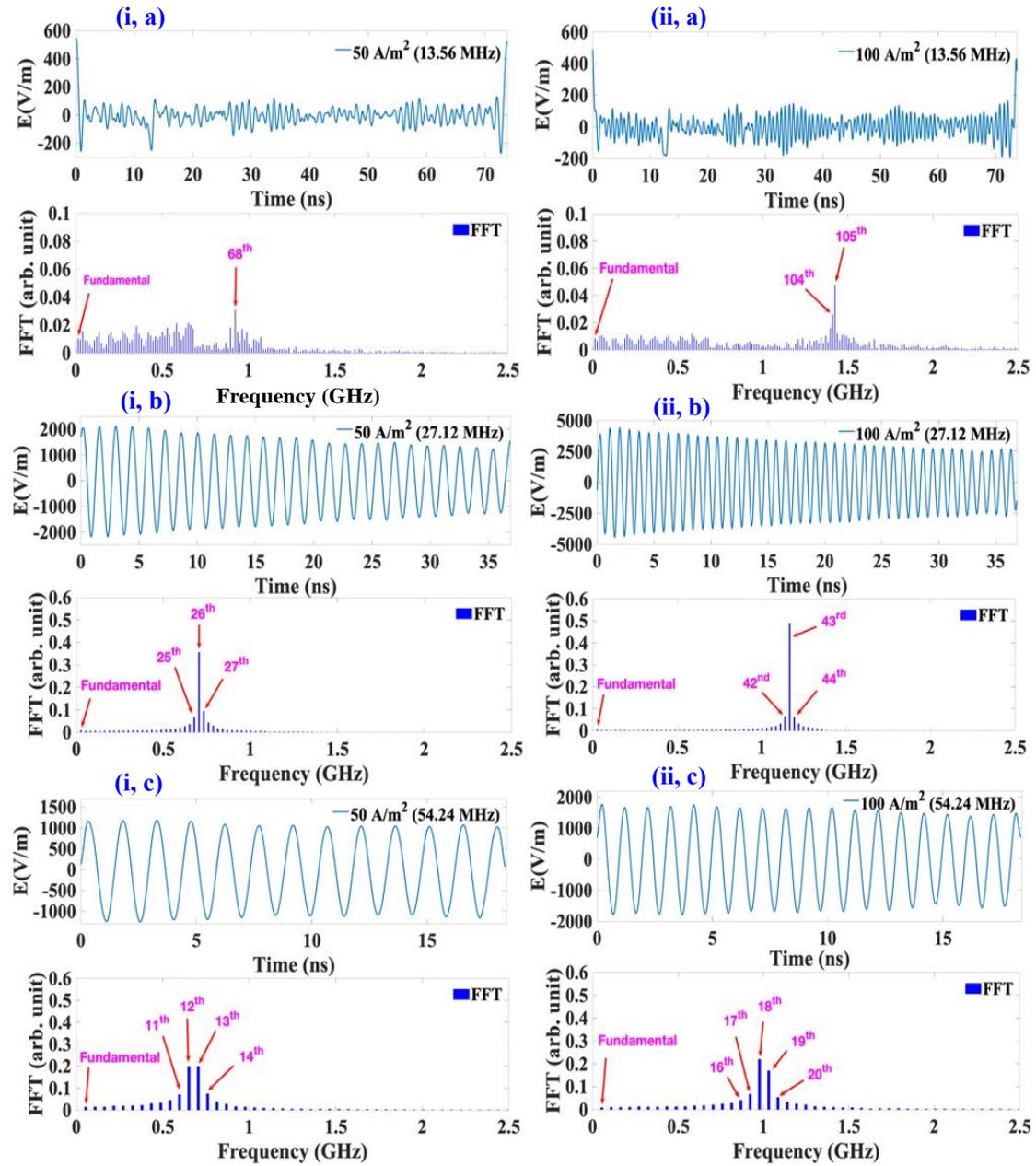

**Figure 6:** Electric field and its FFT at the centre of the discharge for different applied frequencies i.e. 13.56 MHz, 27.12 MHz and 54.24 MHz, and for current densities of 50 A/m$^2$ and 100 A/m$^2$.

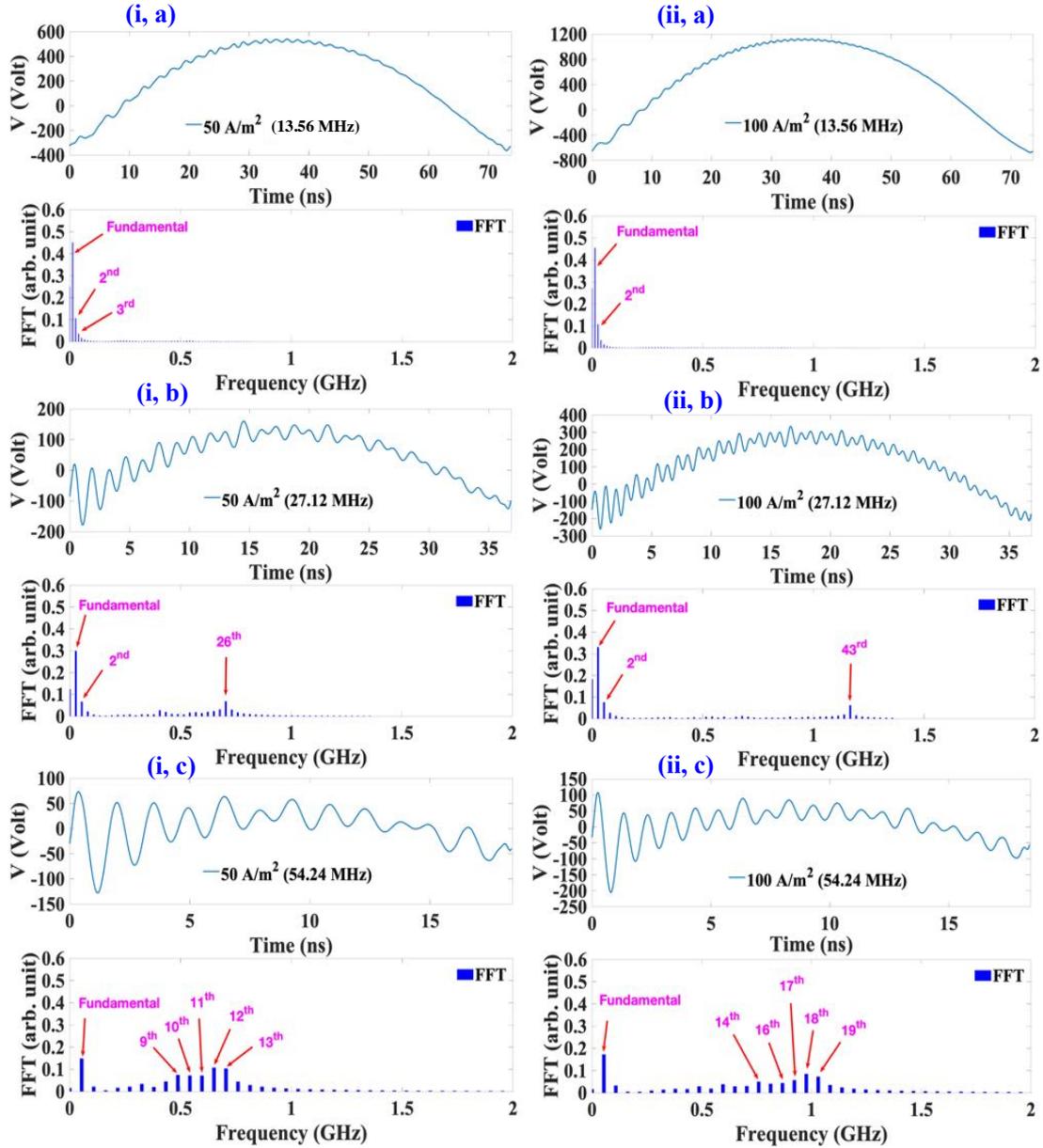

**Figure 7:** Electrode voltage and its FFT at different applied frequencies i.e. 13.56 MHz, 27.12 MHz and 54.24 MHz for current densities of 50 A/m$^2$ and 100 A/m$^2$.

We next present the phase-averaged electron heating in the discharge. This is shown in figure 8. As expected, an asymmetry is observed in the phase-averaged electron heating i.e. it is higher near to the grounded sheath in comparison to powered electrode. The difference between energy losses at the powered electrode and grounded is decreasing with driving frequency suggesting that the discharge is becoming nearly symmetric at higher driving frequencies. When comparing different driving frequencies at 50 A/m$^2$ current density, the heating near to the grounded sheath is decreasing from ~100 W/m$^2$ at 13.56 MHz to ~15 W/m$^2$ at 54.24 MHz driving frequency. At 13.56 MHz, the heating in maximum near to the sheath edge followed by electron cooling in the bulk plasma. A strong negative <*J.E*> is observed near to the sheath edge, which is due to the reflected high energy electrons interacting with the bulk plasma causing net cooling in this region [66]. At 27.12 MHz and 54.24 MHz, the electron heating is further propagating the bulk plasma. An alternate electron heating and cooling i.e. beats like structure are observed up to the centre of the discharge. This is a further evidence of multiple burst of high energy electrons produced from near to the expanding sheath edge. These high energy electron bursts travel through the bulk plasma and produces positive and negative space charge/electric field in the spatial directions, which is seen in the spatio-temporal electric field plotted in figure 4. Since the current must be conserved, the alternated sign of electric field generates positive and negative heating in the bulk plasma. As shown in the inset figures 8 (a) and 8 (b), the corresponding amplitude is higher at 54.24 MHz when compared to 27.17 MHz. This is due to the faster sheath expansion at 54.24 MHz that drives more energetic electron bursts in comparison to 27.12 MHz and therefore one might expect to see an increase in the amplitude of electron heating and cooling at higher driving frequency. Similar trend is observed at a current density of 100 A/m$^2$, except, the phase-averaged electron heating in higher and the amplitude of alternate positive and negative heating is greater in comparison to 50 A/m$^2$. The above results suggest that due to high frequency modulation and multiple bursts like high energy electrons, the heating in these discharges will not be described accurately by simple analytical models.

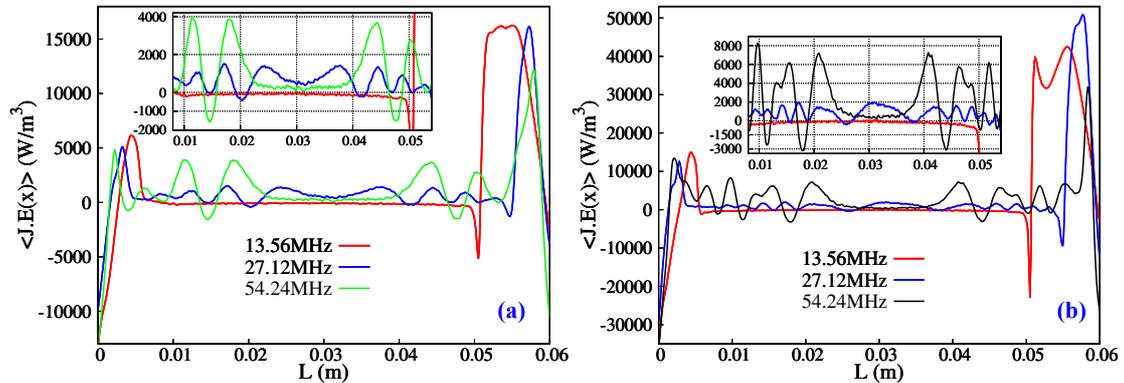

**Figure 8:** Time average electron heating (<J.E>) in the discharge at different applied frequencies i.e. 13.56 MHz, 27.12 MHz and 54.24 MHz for current densities of a) 50 A/m$^2$ and b) 100 A/m$^2$.

## 4. Summary and Conclusion

The effect of driving frequency on discharge asymmetry, electric field/voltage non-linearity and electron heating mechanisms of a low pressure capacitively coupled argon plasma discharge excited by sawtooth like current waveform is investigated using particle-in-cell (PIC)/Monte Carlo Collisions (MCC) simulation. For a constant current density, the plasma density is highest at lower driving frequency and discharge is highly asymmetric. As driving frequency increase, the plasma density decreases, and discharge becomes nearly symmetric. The corresponding ionization rate along with the excitation rates are highest/asymmetric at lower driving and decrease/becomes symmetric as driving frequency increase. Multiple ionization beams are observed from near to the expanding phase of the sheath edge at the grounded electrode. These multiple ionization beams are extending up to the opposite sheath edge and, are higher/dense at lower driving frequency and becoming lower/less dense with a further rise in driving frequency.

An investigation of spatio-temporal electric field in the discharge shows high frequency modulation on the instantaneous sheath edge position at the grounded electrode. The frequency of sheath modulation is highest at lower driving frequency and therefore drive enhanced plasma density and lower electron temperature in the discharge. As driving frequency increase, the frequency of sheath modulation decreases and the transients are more energetic due to higher sheath velocity, reaching up to the opposite sheath edge and modifying its instantaneous position. The Fourier spectrum of electric field at the centre of the discharge confirms the presence of higher harmonics corresponds to the high frequency sheath modulation. Similar non-linearities/higher harmonics are generated on the voltage waveform at the powered electrode. The simulation results are consistent at different current densities. Therefore, it is concluded that the high frequency oscillations drive plasma density and the sheath velocity drive energetic electric field transients in the discharge.

The phase-averaged electron heating near to the sheath edge at both electrodes decreases with an increase in driving frequency. An enhanced high frequency modulation on the grounded sheath at lower driving frequency is responsible for greater phase-averaged electron heating. The spatial distribution of electron heating shows that at lower driving frequency, the phase-averaged electron heating is maximum near to the sheath edge followed by electron cooling in the bulk plasma. However, at higher driving frequencies, along with the electron heating near to the sheath edge, alternate electron heating and cooling is observed in the bulk plasma. The amplitude of such heating and cooling is higher at 54.24 MHz when compared to 27.12 MHz and attributed to spatial positive and negative electric field/charge separation caused by the bursts of high energy electrons produced from near to the sheath edge.